# Wearable µBrain: Fabric Based-Spiking Neural Network

**Frances Cleary[1,2*], Witawas Srisa-an[3], Beatriz Gil [1], Jaideep Kesavan[1], Tobias Engel[1], David C. Henshall[1,4], Sasitharan Balasubramaniam [3]**

[1]Physiology & Medical Physics, RCSI University of Medicine & Health Sciences, Dublin, Ireland

[2] Walton Institute, Waterford Institute of Technology, Waterford, Ireland

[3]University of Nebraska, Lincoln, United States

[4]FutureNeuro, The SFI Research Centre for Chronic and Rare Neurological Diseases, RCSI University of Medicine and Health Sciences, Dublin, Ireland

**\* Correspondence:**
Corresponding Author
francescleary20@rcsi.ie

**Abstract**

On-garment intelligence influenced by artificial neural networks and neuromorphic computing is emerging as a research direction in the E-textile sector. In particular, bio-inspired Spiking Neural Networks (SNN) mimicking the workings of the brain show promise in recent ICT research applications. Taking such technological advancements and new research directions driving forward the next generation of E-textiles and smart materials, we present a wearable µBrain capable of event-driven artificial spiking neural network computation in a fabric based environment. We demonstrate a wearable µBrain SNN prototype with multi-layer computation, enabling scalability and flexibility in terms of modifications for hidden layers to be augmented to the network. The wearable µBrain provides a low size, weight and power (SWaP) artificial on-garment intelligent wearable solution with embedded functionality enabling offline adaptive learning through the provision of interchangeable resistor synaptic weightings. The prototype has been evaluated for fault tolerance, where we have determine the robustness of the circuit when certain parts are damaged. Validations were also conducted for movements to determine if the circuit can still perform accurate computation. As one potential application, we have interfaced the wearable µBrain to live neurons to demonstrate how pressure sensors on the surface can lead to stimulation of neurons. This demonstrates the potential of the wearable µBrain that can be interfaced directly to the brain in the future to stimulate neural circuits due to patients suffering from neurodegenerative diseases, such as stroke or nerve compression that have resulted in the loss of physical sensation in a patient's extremities (lower arm).

# 1    Introduction

Bio-inspired neural networks bring together the convergence of neuroscience, computer science and engineering to create artificial intelligence (AI) that can support applications in human health (Ser et al., 2019). Principles of AI are governed by theoretical neuroscience that is combined with biophysics of the brain to explore novel models for advancing AI, and this is pursued through artificial neural networks (ANN). While diverse applications exist for AI (e.g., autonomous vehicles), applications in health care are increasingly common. This includes applications of AI in medical image processing as well as hardware for wearable sensors (Ker et al., 2017). However, numerous challenges exist in AI and this includes high power consumption, low computing capability and speed of transmission (Covi et al., 2021). These challenges are core to the next generation of AI hardware, and in particular for *Neuromorphic* processors (Indiveri, 2021).

Taking inspiration from the working functions of the human brain and the nervous system, emerging neuromorphic computing architectures are attempting to replicate more efficient brain-inspired neural networks to develop realistic SNN that emulates the manner in which a neuron fires in a biological neuronal network. Neuromorphic chips such as Intel's Loihi (Davies et al., 2018)(A.L et al., 2018), lava software framework and IBMs TrueNorth (Cheng et al., 2017) have driven advancements in this area. Intel's Loihi2 chip facilitates the capability to house one million "neurons" in a chip half the size of its predecessor Loihi1 with *131,072* "neurons" (Davis et al). Such hardware developments are producing new architectural designs and methods enabling scalability and enhanced performance, while reducing energy and cost. From a SNN architectural design perspective, neuromorphic chips are utilizing intelligent memory enabled resistor components, also known as *memristors* (Mehonic et al., 2020), along with the cross bar array design (Innocent et al., 2021) to mimic a biological neuronal network structure. Currently, there is a lack of dynamic SNN devices. Ongoing research is investigating the development of large scale memristive arrays where such cross bar array structures adopt a 2D, vertically 3D layered or stacked architectural design allowing for efficient flow of data between each layer (Covi et al., 2021).

Neuromorphic computing provides a level of advanced AI, such as learning from limited data and adaptive architecture, while also providing computations with low latency. (Stuijt et al., 2021) detailed a µbrain CMOS digital chip that offers ultra-low power and always-on neuromorphic computing that provides reduced latency and energy consumption. Applications using neuromorphic computing are still in their infancy but such devices are envisioned to impact multiple sectors such as biomedical applications, smart neuromorphic biosensors and point-of-care diagnostics (Tuchman et al., 2020). The healthcare sector could benefit from enhanced drug delivery systems (Jayaraman 2020), intelligent prosthetics using flexible neuromorphic electronics (Park, Lee et al., 2020), more efficient data use of image processing tools and technologies driving the next generation of neuromorphic vision sensors (Liao et al., 2021).

The smart materials and E-textiles sector have recently provided new forms of patient monitoring, but such advancements have not been developed towards integrating AI for on-garment intelligence. Healthcare based E-textile solutions currently focus on technologies to remotely monitor patients in their home environment allowing for early intervention. How E-textiles can support such real time and remote unobtrusive monitoring in order to assess the state of the wearer's condition and also to effectively manage their health is an important challenge that needs to be met by smart garment and novel applications in this space. There has been advancements using artificial neural networks across many healthcare applications (Shahid et al., 2019). (Mokhlespour et al., 2019) focused on the classification of physical activities using '*smart garments*', where they validated the accuracy of the smart textile system that featured a smart undershirt and smart socks to assess their accuracy in several postures and activity related movements. In order to classify the various types of physical activities, a number of classification models were utilised and this includes K-nearest neighbour (K-NN), linear discriminant analysis (LDA), and ANN. In recent years, SNN healthcare applications are emerging as this technology advances. (Pang et al., 2020) completed a feasibility analysis into the adaptation of SNN for a structural health monitoring (SHM) system and (Garain et al., 2021) investigated the simulation of SNN and its application for classification of chest CT scan images for COVID and non-COVID cases. Such advancements are still very much standalone applications. Within the E-textile domain, more emphasis and research into how such on-garment intelligence can be embedded into a fabric-based environment needs to be completed.

*Condition-specific* smart garments are becoming more important to patients who suffer from chronic or life threating illnesses and who are required to be monitored or self-monitor their condition on a daily basis. Neuromorphic computing and SNN have the potential to deliver new innovative solutions to meet such needs in a textile based smart garment environment. Wearable artificial neural network computing is gaining traction with research emerging that is investigating the development of a 1D multi-synaptic device to support neural networks in an E-textile environment (Ham, Kang et al. 2020). Wearable memories and computing research as well as applications are focussing on the possibility of embedded and integrated electronic systems in our everyday clothing that interacts with our on-body sensors, near-body mobile devices and even our brain (Rajan et al., 2018).

This paper introduces a brain-inspired wearable μBrain adopting elements of neuromorphic computing architecture and design, to implement a SNN in a textile environment. The wearable μBrain architecture incorporates a SNN to emulate a natural network existing in a biological brain, where each neuron can fire independently of the others by sending pulsed signals to other neurons. It is inspired by the key elements of neuromorphic computing, and this includes 1) the synaptic device, 2) the neuronal circuit 3) the crossbar array and 4) an interconnected communication layered structure used in neuromorphic architecture. Usability is demonstrated through a health focused use case linked to lack of sensation /numbness in the forearm because of nerve compression, a stroke or a neurodegenerative disease. The wearable μBrain processes data inputs coming from the embedded textile pressure sensors in the sleeve textile garment that artificially detect forearm skin sensations based on applied pressure points on the forearm. Despite progress in neuromorphic computing and SNN, there is a lack of current research focused on how such computing advancements can be incorporated into a textile environment enabling a next generation of embedded brain-inspired textile intelligence.

The rest of the paper is organized as follows: Section 2 discusses the materials and methods in detail. Section 3 gives detail around the hardware implementation of the Wearable μBrain. Section 4 gives experimental results and performance insights. Section 5 details application area examples. Section 6 is the final section that concludes the paper and provides insights into future potential work.

## 2    Materials and Methods

Let us now consider the key architectural elements and methods defined in order to progress towards the development of a wearable μBrain. Taking a brain inspired approach, it is important to understand the workings of a biological SNN that will be used to design and develop the wearable μBrain.

Biological neural networks have the capability to process complex inputs, such as electrical impulses, utilizing their structural components consisting of *synapse*, *dendrites*, neuron cell body (*soma*) and also the *axon*. It is estimated that there is 100 billion neurons in a human brain along with a quadrillion synapses (Herculano-Houzel 2009). The neural networks interconnection process involves the dendrites receiving signals from other neurons which in turn is aggregated in the soma and transmitted by the axon to other cells. As neurons become electrically excitable, the action potential electrical pulses that are propagated in the cell and axon cause depolarization of neighboring cells, resulting in a change in the neighboring cell's electric charge distribution ( Kadir et al., 2018). This in turn causes the neighboring neuron to become engaged through the synapse transmission.

SNN brings the promise of low power consumption and energy efficiency when implementing a neural network, as well as analog computation and event driven processing (Pfeiffer and Pfeil 2018). Such SNN properties along with their more accurate representation of the biological neuronal network, make it favorable for low power efficient computational applications driven by event-based inputs and outputs. This opens up opportunities for neuromorphic computing and SNN design architectures and applications to be linked to the E-textiles and the wearable sensor domain. Such applications enable further investigation around embedded on-garment SNN wearable intelligence providing enhanced healthcare and biomedical solutions with event driven, low power and energy efficient properties, within the healthcare sector. This following section details the core modular artificial biological elements and fabric design-based methods required to create and implement a wearable µBrain application, inspired by state-of-the-art neuromorphic computing and SNN models [Fig 1].

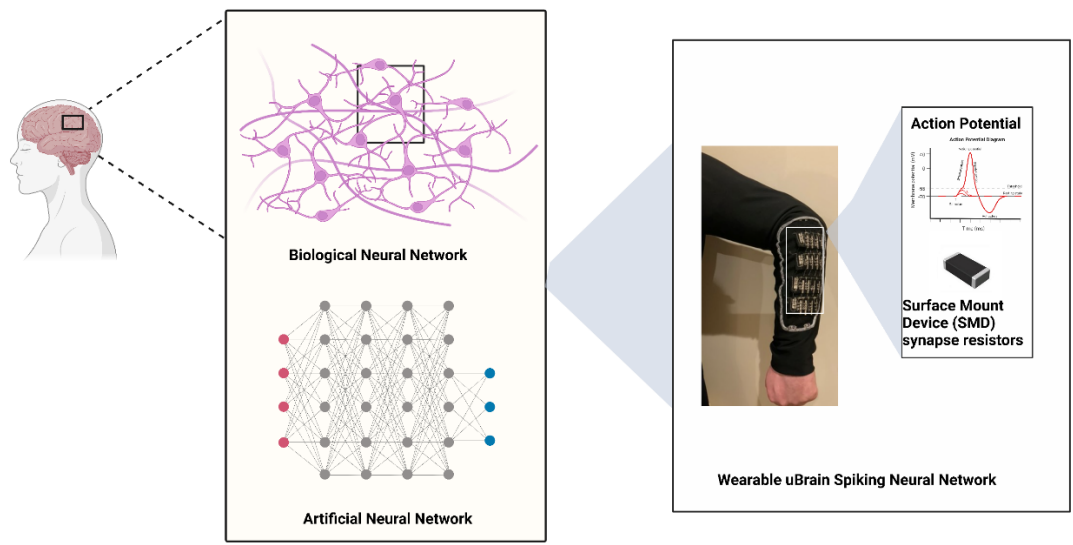

Figure 1 Artificial resistor surface mount device, replicating synapse transmission, core to the SNN concept implmentation in the creation of a wearable µBrain.

## 2.1 Wearable µBrain Architecture

Taking inspiration from the biological neural network, in particular the electrochemical neuronal communications of the brain, here we detail an overview of the modular elements that constitute the wearable µBrain and how these form the overall architecture. Both design features and properties of neuromorphic computing and SNN are advancing and in the wearable µBrain we incorporate elements such as, 1) neuronal circuit, 2) synaptic device, 3) neuromorphic crossbar array, and 4) 3D vertically stacked layered approach. How these elements can be incorporated into a textile environment to support the development of a wearable µBrain SNN architecture will now be discussed further.

### 2.1.1 Artificial Neuron

To artificially mimic a neuron using electronic components we need to replicate its biological functionality (excited state, inhibited state) and integrate this into a wearable fabric-based substrate. The wearable µBrain demonstrates a SNN using a simple neuron replication model limiting the amount of physical electrical components required to be integrated into a fabric environment. To minimize the overall wearable µBrain SNN size in a fabric environment we also use small miniature surface mount device (SMD) components (Molla et al. 2017), which will be elaborated in section 3.

Electrical equivalent circuit models of spiking neurons currently exist (McCulloch-Pitts, Integrate-and-Fire, Hodgkin-Huxley), which contain attributes such as binary inputs, synaptic weights, defined threshold and outputs. We focused on the integrate and fire (IF) (Burkitt 2006) model towards the creation of a wearable µBrain. This has been in existence for many years (Lapicque 1907) and is a simple model using an electronic resistor and capacitor (RC) circuit that incorporates a threshold to mimic the generation of action potential in artificial neurons. The RC circuit is one of the most popular and simple neuronal models as its captures the most important neuronal properties [Fig 2]. It was necessary to create a network of integrate and fire unit neural nodes that could generate a wearable µBrain SNN model that artificially demonstrate neurons' features and functionality. We selected the integrate and fire model to replicate the neuron activity in the wearable µBrain due to the fact that it could integrate synaptic inputs and create a spike when the voltage threshold was reached.

### 2.1.2 Artificial Synapse

The strength of the connection between each node in a SNN and how we incorporate this into the wearable µBrain is a key element of the architecture to ensure we enable interchangeable synaptic weight capability, enabling adaptive learning in the neural network. Depending on the type of SNN, a neural node can have multiple connected synaptic weights. Our defined neural network has one *Fabric Hidden Layer* (FHL) with 3 input synaptic weights to each neural node and a second FHL with 4 input synaptic weights to each neural node [Fig 3]. It is important to include in the overall architecture design the capability to have modifiable weights, as in a biological neural network each neuron can have multiple synapses connected to a neuron all with varying triggering levels that can either excite or inhibit the neural node. To accomplish this in the wearable µBrain, we looked at current state of the art neuromorphic architectural designs linked to SNN.

Neuromorphic computing moves away from the traditional method of separate data storage and computing. Core to the neuromorphic architecture is memory enabled synaptic devices such as the synaptic memristor (Yao et al., 2020) 2) and synaptic transistor (Wan et al., 2020), that both store and compute hence eliminating the *Von Neumann bottleneck* (Park 2020). For the implementation of the SNN, we had to consider the synaptic weights required as connection strengths between the neural network nodes. The synaptic weights and their strength levels play an important role in the SNN as they mimic the firing impact one neuron can potentially have on another neuron (Herzfeld, Beardsley 2011). To simulate the weight connections, we used SMD resistors of varying electrical resistance (*Ohms*) for each input connection to each neural node.

### 2.1.3 Neuromorphic Inspired Crossbar Array Architecture

To implement a fully connected SNN, the neuromorphic crossbar array-like architecture consisting of rows and columns has been used in hardware implementations of ANN's (Xia et al., 2019)(Balaji et al., 2019). Memristor artificial synapse device advancements are also driving the adoption and use of memristive crossbar arrays enabling low power memory computing and highly scalable architectures. (Ji et al., 2016) demonstrated a *1 transistor*, *1 resistor* crossbar array architecture to eliminate undesired cross talk. (Lee et al., 2020) details the use of a hybrid memristor and transistor, known as *memtransistors* that acts as dual gates for the traditional crossbar array sneak current problem. Taking this neuromorphic crossbar array row and column design structure, we have incorporated it into a wearable µBrain structured design, to accommodate a functioning neural network layout with an inbuilt high density of synapses using SMD resistors [Fig 2]. Circuit theory for the crossbar array utilizes a combination of Ohms law and Kirchoff's Voltage law supporting the crossbar array electrical computations.

It is worth noting here the sneak path cross bar array issue which limits accurate sensing. The crossbar array structure creates *sneak path currents* through the array's rows and columns. Various techniques have been investigated and used to try to eliminate this issue, these include *transistor gating* (Yuan et al., 2021), *diode gating* (Seufert et al, 2021), *row grounding technique* (Fouda et al,. 2019), *multistage reading* (Sun et al., 2020). To address this issue in the wearable µBrain cross bar array fabric design we will adopt and use the row grounding technique. Here unselected input rows in the wearable µBrain will be grounded directing sneak path currents to ground. From a component and textile point of view, no additional components are required to implement this solution and hence it suits the design direction, limiting the number of physical components to create the wearable SNN.

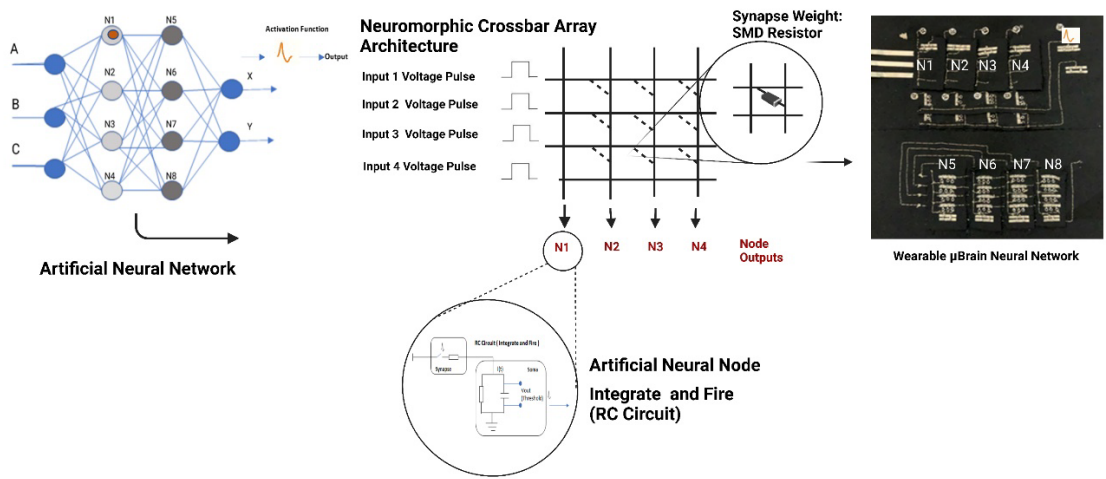

Figure 2 Replication of Neural network neuronal properties using RC electrical circuit (Integrate and Fire Model) and a neuromorphic crossbar array with SMD resistor weights simulating the action potential of a neuron, implemented in a wearable µBrain neural network.

### 2.1.4 3D vertical stacked layered wearable µBrain SNN

The novelty of the wearable µBrain SNN exists in its 3D vertically stacked FHL design structure. 3D integrated system design in neuromorphic computing (An, 2019) [Fig 3] and artificial intelligence hardware applications are generating more sophisticated and scalable systems acting as key enablers driving adoption (Yang et al, 2018) (Koyanagi et al., 2001). Such design flexibility and scalability properties provide a more modular, cost effective and energy efficient (Xia et al., 2019) design methodology suitable for the wearable µBrain. To accommodate this layered and vertically stacked FHL approach, careful consideration was given to 1) the required communication interconnections between the fabric hidden layers, 2) the design and placement of fabric patch artificial neural nodes on the FHL, and 3) the interconnection of rows/columns supporting a crossbar array layout [Fig 3].

Interconnections to support the communications from FHL to FHL*n* using snap connectors were placed at set points parallel to the artificial RC neural nodes, allowing ease of connection and assessment of the voltage output ($V_{out}$) measurements readings from neural nodes in FHL1 to its connecting FHL*n* [Fig 3]. Placement of the artificial spiking neural network nodes on the FHL were split based on their hidden layer location in the neural network structure [Fig 3]. Fabric based patch neural nodes were created to house the crossbar array row/column conductive thread connections and also to house the artificial synaptic SMD resistor weights and RC neural nodes. Each Fabric patch node has the capability to be easily removed, replaced or updated adopting a modular design approach to the overall wearable μBrain all the while accommodating offline adaptive learning through SMD resistor weight selection updates as required.

As part of the multi-layer fabric topology, it is important to include functionality to dynamically interchange the resistor weights that replicate the synapse weight in a biological neural network. Depending on the desired output, a selected SMD resistor weight was connected to the array column conductive thread line that fed into the RC circuit artificial neural node [Fig 3]. The adopted design approach using fabric node patches on the FHL, allowed for ease of interchange of the SMD resistor weightings as desired, using conductive thread stitching and also provided clear structured recognition of the defined neural nodes in the FHL's which supported performance assessment and validation.

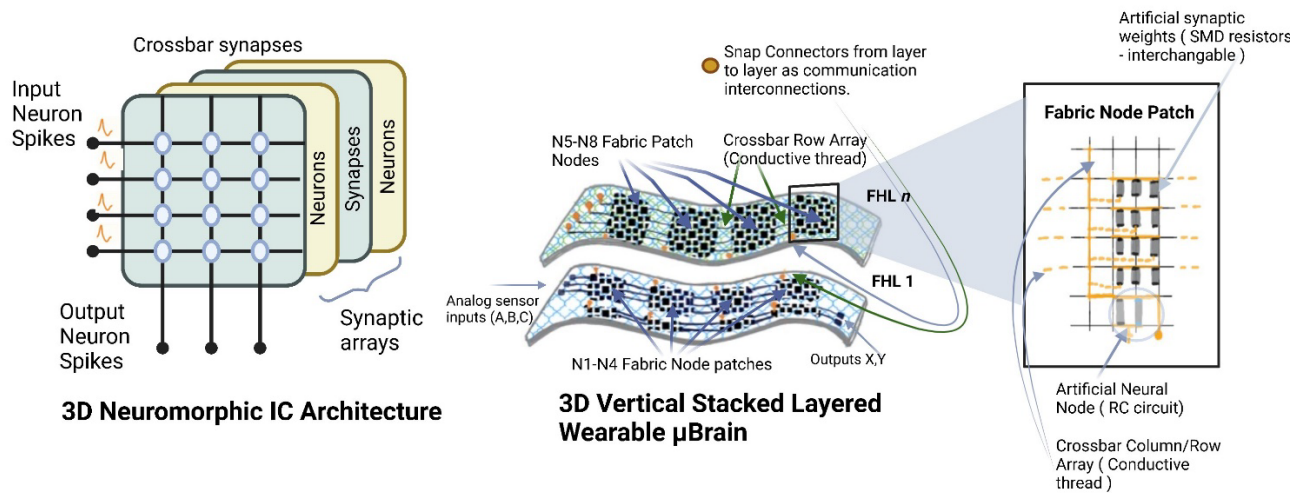

Figure 3  Neuromorphic 3D vertical stacking architecture inspiring the Wearable μBrain 3D vertical fabric layered architectural structure.

Advantages of the 3D vertical stacked fabric layered approach include:

- A modular design methodology enabling scalability and extendibility of the prototype for future reconfiguration requirements.
- Provision of enhanced energy efficiency due to the 3D vertical stacked FHL's with reduced power dissipation.
- Interchangeable node fabric patches in the adopted SNN layered design approach, with offline adaptive weight selection allowing SMD resistors of varying resistance levels to be interchanged depending on the application use case.
- Limited cost to implement in a fabric environment (low-cost fabric, SMD devices and conductive thread) with reduced area size, applicable for varying application-based use cases.

## 2.1.5 Interchangeable SNN Weights

For the majority of SNN, current learning methods base themselves on the adjustment of the synaptic weights. The provision of three optional SMD resistor weights for each neural node in the network (*N1-N8* and *X, Y*) [Fig 4 (a)(b)] allows us to directly implement an offline adaptive interchangeable synaptic weight learning method for the wearable µBrain SNN by selecting our preferred SMD resistor weight. The actual desired output is acquired, through selective adjustment of the weights. For our wearable µBrain SNN we have created a dedicated logic truth table based on 3 inputs and 2 outputs (*X, Y*) [Fig 4(c)]. Adopting this logic truth table, it supported the offline adaptive learning of the wearable µBrain using the iterative adaptive interchangeable weight selection process to acquire the desired outputs (*X, Y*).

The wearable µBrain SNN not only has the flexibility to interchange connected weights to modify the workings and outputs of the SNN, but SMD resistors of varying resistance levels can be added and removed as required through simple replacement and soldering of the required SMD resistors into the patches fabric tape row lines [Fig 4(b)]. Having such an easily modifiable wearable µBrain SNN, makes the prototype extendable and capable of facilitating such changes producing varying output results and hence a variety of truth table use case applications. A basic example of this is demonstrated in [Fig 4(a)] where a change of SMD resistors for output node *X* and *Y* results in a change in the outputs results for inputs *010* and *101* for wearable µBrain SNN which can be see through the desired output change in the logic truth table [Fig 4(c)].

Such flexibility allows for varying desired truth table output options (*X, Y*) to be considered and implemented using the wearable µBrain SNN framework, where a dedicated truth table aligned to a specific health related application can be adopted and implemented. SMD resistor resistance levels that act as node synapse connectors in the FHL 1 and FHL 2 can also be altered to further modify the outputs from the logic truth table in order to alter the wearable µBrain applications and use cases.

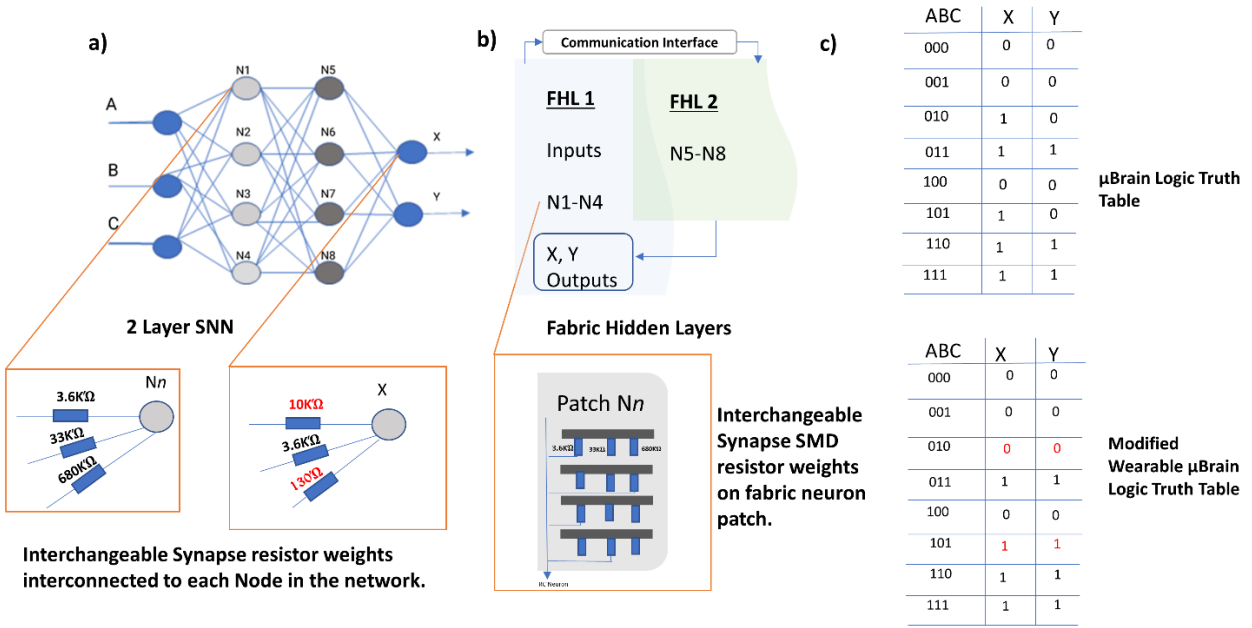

**Figure 4** a) SNN representation of interchangeable selectable resistor weights b)Fabric SNN layouts using a 2 layered FHL approach supporting individual fabric patch nodes with interchangeable weight resistor SMD devices c)Logic truth table applied with desired outputs X,Y

### 2.1.6 Key systematic modules of the Wearable µBrain Architecture

As the IF model RC circuit was used to simulate the neuron properties and the event driven processing functionality of the SNN, an architecture was defined that replicated this electrical circuit functionality in order to build out a working wearable SNN. Key architectural specifications are summarized as follows.

- Use of SMD based RC circuitry simulating the functionality of a neural nodes. These RC node circuits were utilised in the architectural design to implement a network of event driven spiking artificial nodes to create the wearable µBrain SNN.
- The use of SMD resistors as artificial synapse connectors to the neural nodes acting as the required node interconnection weights.
- Each neural node electrical circuit structure to be integrated in a fabric environment using fabric patches to represent nodes.
- Each hidden neural network layer to have its own fabric layered surface area, allowing additional FHL's to be added or removed depending on the application and level of complexity of the wearable µBrain. This replicates a 3D design element, similar to design structures appearing in neuromorphic architectural designs, enabling customizability within the wearable µBrain design.
- Interconnection of the FHL's of the fabric based Wearable SNN using snap connectors, allowing flexibility to easily deconstruct and construct variations of the wearable µbrain SNN. This introduces a high level of flexibility and modularity to the overall architecture design.
- The architecture is designed in a manner to accommodate SWaP AI (low size, weight and power) in a fabric-based application.

## 3     Hardware Implementation

To implement a hardware version of the wearable µBrain SNN, we had to consider 1) the method of incorporating SMD resistor and capacitor components of the defined architecture into a fabric environment 2) Development of a scalable wearable µBrain SNN, with complex hidden multi-layer on-garment intelligence. This section will provide an overview of the prototype hardware implementations completed.

### 3.1    SMD components in Fabric environment.

Taking the defined SNN topology in section 2.1, how to apply this into a fabric environment needed to be considered. In order to reduce the overall size of the fabric based wearable µBrain SNN, surface mount devices were used in the design and construction phase. Adding surface mount devices to a fabric environment was not an easy task, as they are predominantly manufactured for mounting on printed circuit boards and not for embedding in fabric environments. (Molla et al, 2017) present a stitched method of fabricating E-textiles with SMD light emitting diodes (LED) devices. This process to incorporate SMD resistors and capacitors was not viable for our SNN application hence the following process was trialled and used to embed the SMD into a fabric environment.

SMD package type *0805* for the required resistors and capacitors components (*3.6KΩ, 33KΩ, 680KΩ, 220nf*) were selected. *SMD 0805* components have a dimension of (*2.0 x 1.3*) millimetres. To create the synapse resistor weight interconnections for each node, one end of the SMD was soldered to a conductive nylon fabric tape. The nylon conductive fabric tape can be bent and twisted and has an adhesive backing, so this provides the flexibility required in order to add SMDs to the wearable material fabric environment. The corresponding end of the SMD, required the capability to connect to a conductive thread-based circuit connection (Madeira *HC12* Conductive thread, Resistance: <

*100Ω/m*) in order to support the synapse weight SMD resistor selection and implementation of the offline adaptive learning process of the SNN. Here a *3mm* Jump Ring (Silver Plated) was soldered to the end of the SMD components, providing a circular ring attachment to stitch the conductive thread to[Fig 5(b)]. This allowed us to implement a flexible and changeable sewable SMD connection limiting the dependency solely on soldering techniques. For the wearable µBrain, we implemented a FHL neural network containing *8* hidden neural nodes (*N1-N8*) using the IF model RC circuit to replicate the artificial nodes in the SNN.

Depending on the number of nodes in a FHL, a fabric node patch-based approach was incorporated into the design where a single fabric patch represented a single neural node and its interconnecting SMD resistor weights. Multiple patches positioned in parallel on the FHL created the hidden neural node components for the selected FHL [Fig 5(c)].

To implement the RC neural node components, *2* parallel lines of nylon conductive fabric tape were used and the SMD devices were soldered directly to the fabric tape to create the connections [Fig 5(a)].

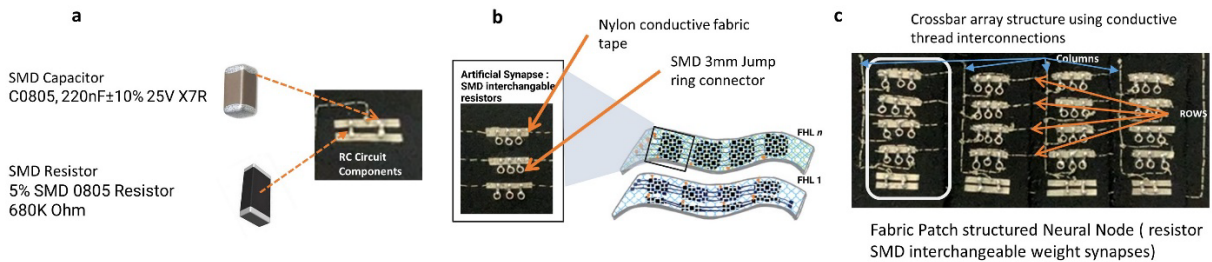

**Figure 5 a) SMD resistor and capacitor components to create Neural Node RC circuit. b) Interchangeable (3.6KOhm, 33kOhm, 680kOhm) SMD resistor weights to represent artificial synapse c) FHL design structure and fabric node patch crossbar array architecture implementation**

Once the RC node component elements were prepared separately, they were then secured to the FHL surface area material and interconnections completed using conductive thread. The conductive thread based interconnections implemented the cross-bar array architecture with defined rows and columns.[Fig 5(c)] . Ten RC node components (*N1-N8*, *X*, *Y*) were created along with *108* individual SMD resistors and *8* fabric node patches. The individual modules were then amalgamated together to form the wearable µBrain [Fig 5,6,7]. A threshold of *2.3 Volts* was set, determining the state of the output logic (<2.3Volts = '0', ≥2.3Volts = '1'). During validation an input power supply of *5* Volts was utilized to simulate the inputs (*A, B, C*) as required to correspond to the truth table inputs.

### 3.2 Multilayered Spiking Neural Network.

Each hidden neural network layer was placed on a separate FHL during the construction phase with RC circuit-based node fabric patches containing artificial SMD resistor weight synapses. Network layer *1* (*N1* to *N4*) on *FHL1* and SNN Hidden layer *2* (*N5* to *N8*) on *FHL2* [Fig 6 (a)(b)], adopted a *3D vertically stacked* layered architectural fabric design approach, that could be scaled as required to include further fabric hidden neural network layers.

This FHL fabric based layered design approach enables each neural network hidden layer to be easily detached, reconnected or a newly modified additional hidden layer, with varying interconnected synapse weights to be incorporated. The neuromorphic crossbar array design was incorporated, using conductive thread to stitch the interconnections (rows and columns). On each fabric node patch a vertical conductive thread line was stitched enabling interconnection of the desired SMD resistor weights to the RC circuit neural node [Fig 6(a)]. Snap fasteners (*7mm*) were used to support the communication interface from *FHL1* to *FHL2* [Fig 6(b)]. The ability to easily manipulate the fabric environment from a flat 2-dimensional fabric layer to a stacked 3D vertical multilayer, with hidden NN layering dimensionality, enhances the functionality and scalability of the wearable µBrain SNN allowing a flexible and dynamic hidden NN layer interconnection capability.

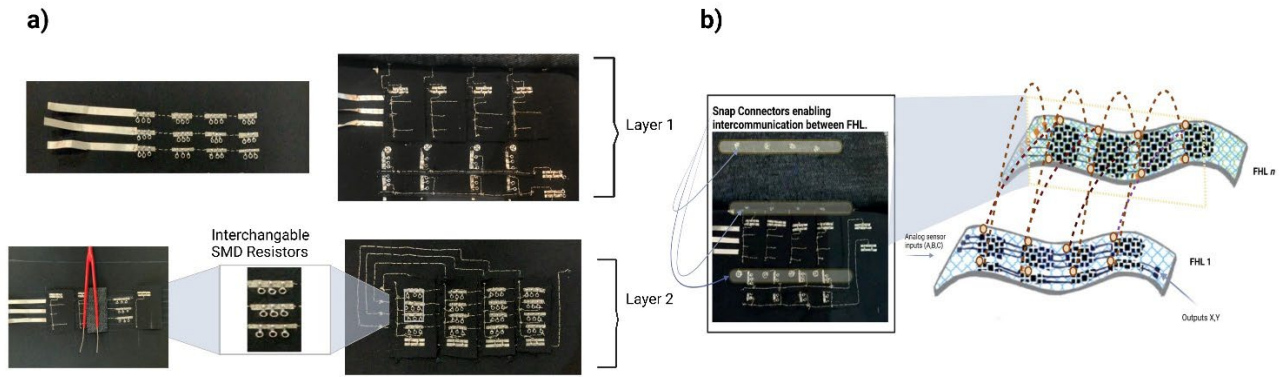

Figure 6 Wearable µBrain Implementation Overview

Following completion of the multilayered prototype, it was decided that placement of the wearable µBrain would be located on the arm of a garments sleeve. Two possible placement options 1) upper arm 2) lower arm were considered [Fig 7]. The wearable µBrain with an approximate size of *15cm* in length and *11cm* in width was sewn onto a sleeve made from lycra fabric. An opening at the front of the sleeves arm location was included allowing the top layer of nodes (*N5* to *N8* and output *X, Y*) to appear through, allowing measurements to be obtained during the validation testing phase.

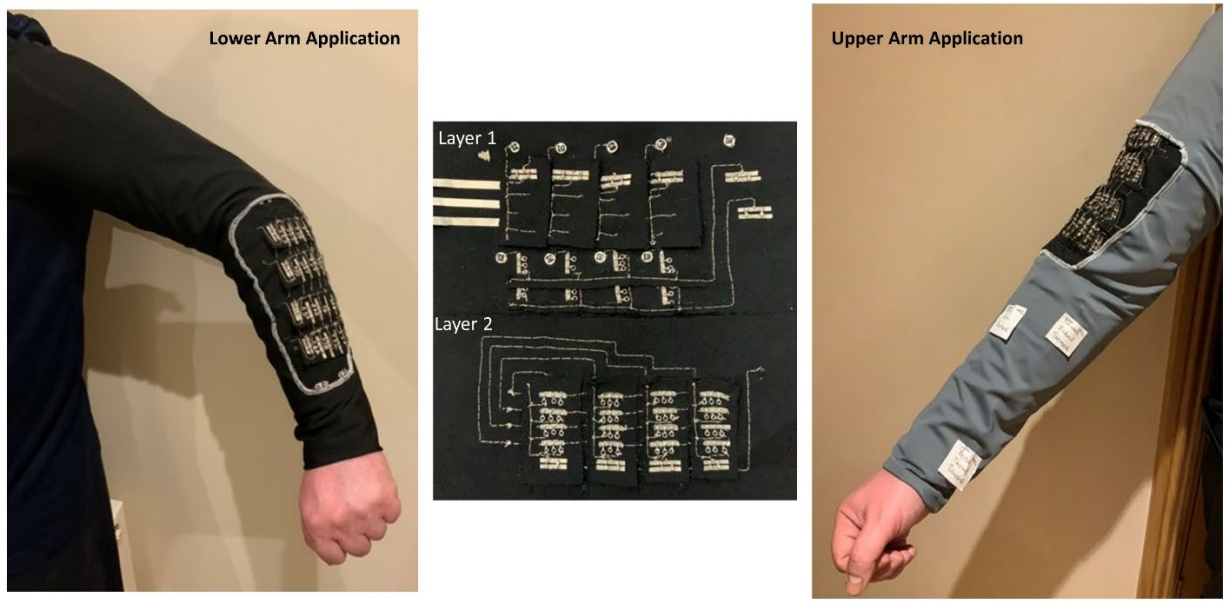

Figure 7 Prototype Implementation of a Multi-layered Wearable µBrain

## 4 Results

This section presents the results obtained from the experiments focusing on the operational functionality of the wearable µBrain SNN on a sleeve and observing and comparing the actual wearable neural network outputs with the expected outputs [Fig 4 (a)(c)]. Utilizing the *Integrate and Fire* model in the wearable µBrain neural network to replicate the functional dynamics of a spiking neuron regulated by resistor synaptic weights [Fig 2], we will validate the offline adaptive learning approach adopted through manipulation of the weights to decrease the errors to achieve the desired output. The tests performed on the wearable µBrain *fully connected* multiple hidden layer SNN network, includes (1) Fault tolerance testing, and (2) Wearability and usability performance results.

### 4.1 Fault Tolerance Testing

As with any wearable or fabric-based material, it is open to the electronic elements and, hence, the possibility of fraying, ripping or damage to the fabric environment needs to be considered. As part of the testing, we investigated wear and tear from a fault tolerance perspective. Taking into consideration the design and layout of the wearable µBrain, we divided the visible surface area up into fault testing zones (*Zone 1* to *Zone 4*), as shown in [Fig 8]. Fault tolerance tests were simulated through the disconnection of specific conductive thread links located on the border area of the fabric node patches, in order to fully observe the potential impact if such fraying or ripping was to occur.

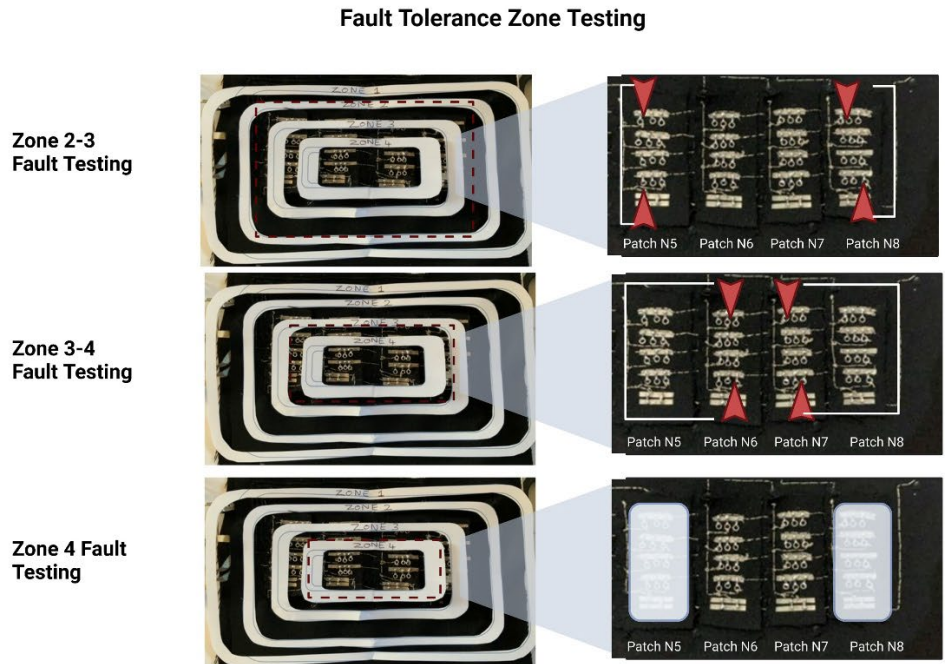

**Figure 8** Overview of fault tolerance zones applied ranging from zone 1 to zone 4 with the corresponding key pinpoints where potential fraying could occur in order to simulate and observe the impact on the functionality of the prototype.

Observations from the simulated fault tolerance tests shown in [Fig 8, 9(a)] can be summarized as follows

- Patch *N5*: *Row 1* SMD resistor disconnected: For input '*010*' when SMD resistor *3.6KΩ* was disconnected, this demonstrated a decrease of approximately *16 -17%* in the voltage output reading for both *X, Y* outputs, whereas for input '*101*' there was an increase of around *14%* to *24 %* in *X, Y* outputs. Patch *N5* demonstrated an increase of *65%* to *67%* for input *001* and *101*, while it demonstrated a decrease of *27%* to *32%* across inputs *010* and *011*.

- Patch *N5*: *Row 4* SMD resistor disconnected: Disconnecting *680KΩ* SMD resistor on *Row 4* Patch *N5* resulted in a decrease of *28%* voltage across patch *N5* input *100*. This also led to an increase in the range of *7* to *9%* for patch *N6* and *N7* inputs (*001,011* and *110*).
- Patch *N5*: *Row 1* SMD and *Row 4* SMD disconnected: Disconnecting both the SMD resistors *3.6KΩ (row 1)* and the *680KΩ (row 4)* on patch *N5*, resulted in a similar output result to just removing the *3.6KΩ* resistor, but with slight differences in the output voltage readings.
- Patch *N8*: *Row 1* SMD resistor disconnected: When disconnecting patch *N8*, it showed a *25%* increase in output *X (101)* and a *27%* decrease in output *Y* for input (*011*). Patch *N8* showed fluctuation increase of *38-39%* for inputs (*001,101*) and showed a decrease of *21 to 23%* for inputs *010* and *011*.
- Patch *N8*: *Row 4* SMD resistor disconnected: Upon disconnecting the *680KΩ* SMD resistor from *Row 4* patch *N8*, this resulted in an increase to input *001* by *31%* and a decrease to its corresponding *N7* by *37%*. This in turn led to an increase in output *001 X* output by *28%* and *101* by *18%*. Input *010* demonstrated a decrease across both *X, Y* outputs between *11* to *16%*.
- Patch *N8*: *Row 1* and *Row 4* disconnected: Disconnecting both resistors, demonstrated a clear increase in *001*, *100* and *101* output readings coming from *X* within the range of *25-34%*. Decrease for output *Y* were displayed (range *14-25%)* for inputs *010* and *011*. Disconnection showed a spike increase of *50-58%* for both *N5* and *N8* (input *001*), in contrast a decrease of *51%* can be seen for *N7*(input *101*).

In general, when these simulated fault tolerance use cases on the wearable µBrain prototype were completed, the core operational functionality of the prototype remained mostly intact. The current redistribution amongst the nodes in the SNN impacted mainly the output readings obtained from the triggered inputs *010* and *101*. The main reason here was due to fluctuation of the output *X, Y* results around the set threshold level of *2.3 Volts*, following the fault tolerance connection modifications. Overall good functionality remained intact with limited impact on the expected outputs results and the workability of the prototype.

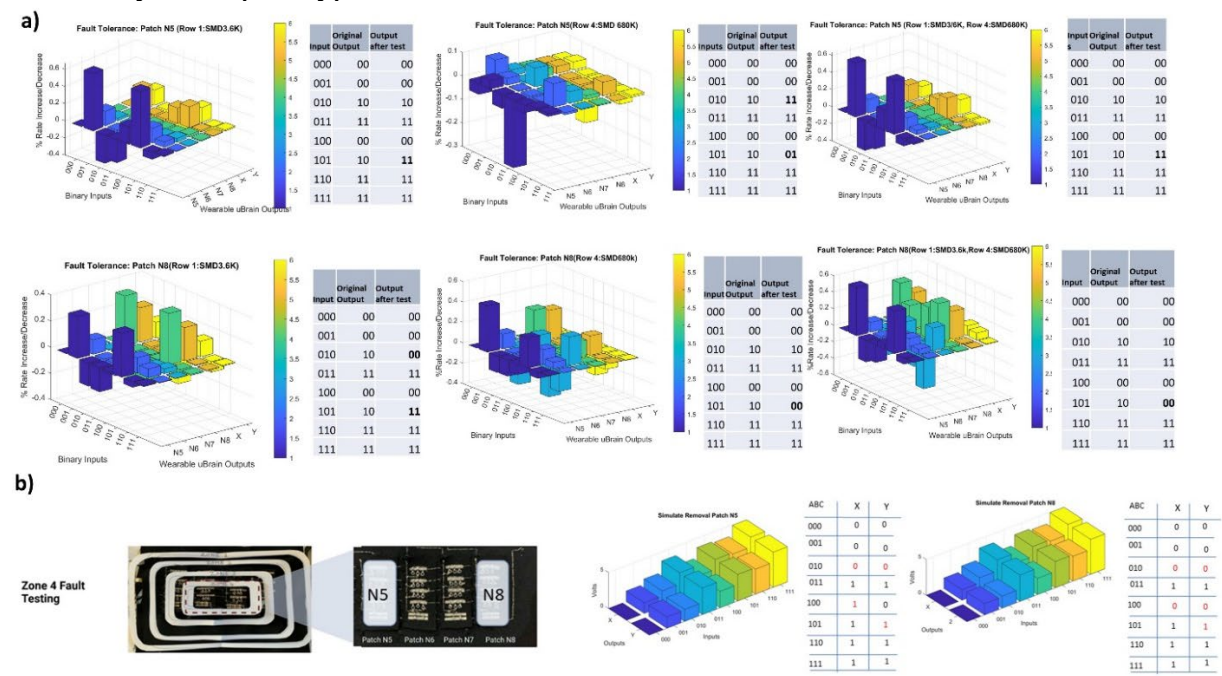

**Figure 1 a)Fault tolerance results simulating potential impact of fabric fraying across Zone 2 to Zone 4. b) Fault tolerance results simulating disconnection of patch for Node 5 and Node 8.**

To demonstrate and test the impact of removing the dedicated fabric node patches has on the prototype, we focused on the disconnection of 1) Patch *N5*, and 2) Patch *N8* in zone 4 [Fig 9(b)]. In the prototype we could easily simulate the failure of these nodes through the disconnection of the communication layer snap connector coming from *FHL1* to *FHL2*. We first disconnected patch *N5* and tested the resulting output voltage readings and secondly, we then disconnected Patch *N8* and again tested the output voltage readings coming from *X, Y* on the prototype.

With the defined set threshold of *2.3 Volts*, we could then complete a corresponding truth table based on the obtained output results and map this to the original truth table to assess the impact that loss of certain node functionality would have on the prototype. [Fig 9(b)] conveys the output voltage readings obtained following the disconnections. We also present the corresponding truth tables, highlighting the changes incurred following the disconnection. It was observed from the results that the truth table outputs for inputs *010,100,101* were altered, because of the fabric node patch disconnections altering the desired outputs expected for the SNN. Such a drastic alterations to the prototype through the complete lack of functionality of a neural node (in our case *N5* and *N8*) did have more of a negative operational impact on the working of the prototype, but this was expected as removal or disfunction of a neural node in a SNN will alter the desired and expected output.

### 4.2 Wearability and Usability performance results

E-Textiles and smart garments often adopt a user centric design approach ensuring relevant end user usage scenarios are considered. Performance assessment and the reaction of the prototype under such user centered scenario constraints need to be considered. Based on the placement location of the wearable µBrain on the lower arm, we specifically focused on the analysis of lower arm movement assessing the impact on the wearable µBrain sleeve prototype and the output results obtained. Such performance and usability tests involved movement of the arm in a walk-like motion while the sleeve with the prototype was worn by an individual.

To test the level of endurance the prototype had under such motion-based conditions, we increased the intensity level of the arm walking movement motion from normal pace to fast pace and obtained output voltage measurements from *X, Y* on the wearable µBrain. The following use cases were trialed and tested: 1) normal - no arm motion, 2) 6 walking iteration arm movements at normal pace, and 3) 32 walking iteration arm movements at a fast pace. Data readings coming from these use cases allowed for cross comparison against the expected output results when the wearable prototype was in a static state. Setup required the wearable µBrain to be added to the fabric sleeve and the sleeve worn by an individual to simulate the walking paced arm movement. The outputs *X, Y* from the wearable µBrain prototype were connected to oscilloscope probes allowing the max ($V_{max}$, the highest value in the waveform display), min ($V_{min}$, the lowest value in the waveform display) and mean (the arithmetic over the entire waveform) voltage values of the output results to be displayed, saved and visualised on the oscilloscope. An accelerometer was attached on the side of the arm to track the arm movement iterations as shown in [Fig 10(a)].

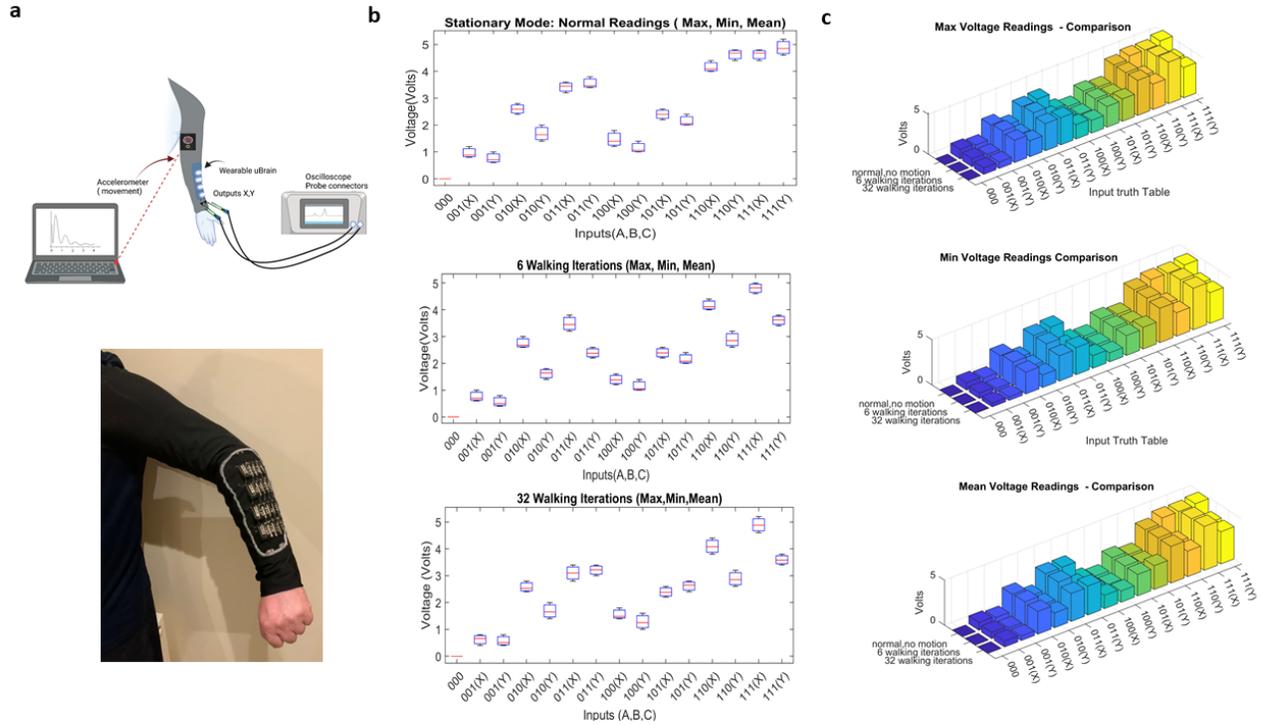

**Figure 2 a) Setup configuration to test the Wearable µBrain while on the sleeve and connected via probes to the oscilloscope, with an accelerometer attached to side of the arm to track arm movement, b) $V_{Max}$, $V_{Min}$, mean boxplots conveying voltage variance for each of the three trial test scenarios, and c) comparison of the output voltages across the three defined scenarios to cross compare all the max, min and the mean levels.**

Resulting output voltage readings were taken at defined intervals mid-way through the selected scenario under test. The max, min, mean values obtained from the oscilloscope measurement readings are represented in boxplot shown in [Fig 10(b)] when varying truth table inputs were applied. Only slight variations were present in the voltage readings obtained when the scenarios were implemented with measurements taken. Fluctuations could be seen in the output voltage for *X, Y* when the inputs *110,111* and *011* were applied.

- For Input *010* (*X*), taking the mean output voltage reading for a stationary scenario produced an output voltage of *2.59 Volts*. Readings obtained from 6 walking iterations to 32 walking iterations showed a slight decrease from *2.68 Volts* to *2.53 Volts*.
- For input *101(Y)*, a slight increase in the mean voltage reading from *2.08 Volts* (6 iteration walking) to *2.65 Volts* (32 iteration waking) was observed. In general, the observed and measured readings during the movement iterations did not impact negatively on the operational functionality of the wearable µBrain.

A cross comparison of the $V_{max}$ readings across the three scenarios highlights a slight decrease of ~ *0.2 Volts* between normal and 32 walking iterations scenarios for inputs *011* (*X*) and *011* (*Y*). A cross comparison of the $V_{min}$ demonstrated a drop in voltage from *4 Volts* to *3.8 Volts* for input *110* (*X*) and a decrease from *4.4 Volts* to *2.6 Volts* for input *110* (*Y*).Overall, the most noticeable difference was the output voltage readings taken for *110* (*Y*), *111* (*Y*), *011* (*Y*) where slight voltage drops occurred for the 6 and 32 fast paced walk iteration results, as shown in [Fig 10(c)].

Based on the fault tolerance, wearability and usability testing performed on the wearable μBrain on a fabric sleeve environment, the outcome following the tests was of an acceptable level where the impact of the tests did not greatly alter the operational functionality of the prototype. It was noted that the threshold level set for the overall wearable μBrain architecture, was an important element to consider, as fluctuations seen in the operational level of the prototype mainly impacted output voltage readings that were within this set threshold level (e.g. *010*, *101*), as observed measurements randomly shifted above and below this set threshold voltage value.

### 4.3 Wearable μBrain Applications

Health conditions such as strokes, neurodegenerative diseases or even nerve compression can cause numbness or loss of sensation in parts of the human body (Jang,Yeo., 2017) (Balaji 2021). Numbness is where an individual cannot feel pain or even light touch. Lack of temperature along with lack of sense of positioning of the arm can impact the users' balance and coordination. The capability to feel sensations is directly linked to the skin receptors sensory nerve fibers that pass detected signals through the spinal cord into the brain stem and to the cerebrum in the brain, where the signals are interpreted and processed as shown in [Fig 11(a)] and [Fig 13(c)].

### 4.3.1 Wearable μBrain Forearm touch sensation simulation Application

To demonstrate an application of the wearable μBrain, we focused on a use case where there has been compression or damage to the lateral cutaneous nerve in the forearm (Wu et al, 2019). This nerve is responsible for the skin sensation and touch receptors in the forearm. As a result of the nerve compression and damage, the patient suffers from numbness in the forearm. The wearable μBrain acts as an intelligent integrated garment based SNN that can classify haptic sensing coming from artificial pressure sensors embedded into a garments sleeve aligned to the lower elbow, mid forearm and base of thumb. This enables embedded computation through the wearable SNN to decipher which part of the forearm has been touched with the end conceptual goal of relaying an alert (based on the sensing triggered data) back to the cerebrum in the brain for processing.

The artificial pressure sensors were constructed using a combination of *2* layers of plain fabric, *2* layers of conductive fabric (*EeonTex NW170-PI* fabric) and *2* layers of velostat, which is a pressure sensitive conductive material that reduces in resistance once pressure is applied. The textile pressure sensors were stitched into the lower part of the garments sleeve. Their exact location and placement in the sleeve matched to the sensing points that aligned to the lateral cutaneous nerve, as shown in [Fig 11(a)]. Tests were completed to observe the output voltage increase, when pressure was applied to the sensors based on the defined truth table input combinations. A standard *5 Volts* power supply was used as an active input when required based on the truth table triggered inputs.

[Fig 12] demonstrates the output voltage increase once the corresponding input *A*, *B*, and *C* sensors were pressed. When a touch pressure was applied to the textile pressure sensor, the resistance in the velostat fabric pressure sensor decreased. This caused an increase in the voltage output from the sensors, which simulated a touch sensation. This voltage range acted as a trigger for the wearable μBrain inputs where the embedded SNN completed the on-garment intelligent computation. The output voltage readings obtained from the wearable μBrain were recorded and cross compared with the desired range levels of the defined truth table. Output voltage readings from nodes *N5* to *N8* and output *X*, *Y* are displayed in [Fig 12], highlighting the increase from pre to post output voltage ranges once the varying input sensor combination (*A*, *B*, *C*) on the sleeve were selected and pressed.

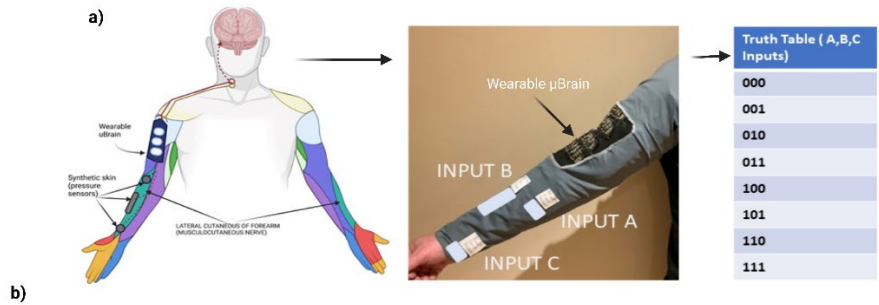

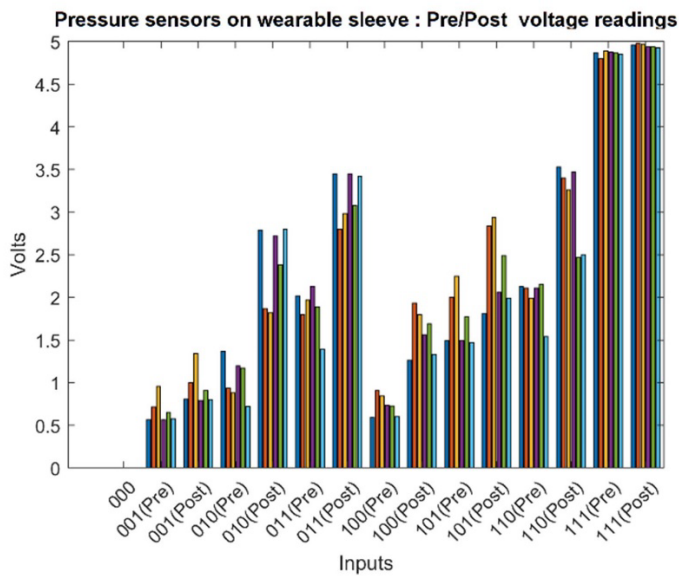

Figure 11 a) Haptic sensing communication flow from the skin to the brain via the spinal cord and Implementation of synthetic pressure sensors and mapping to truth table inputs. b) Sensation touch scenarios to validate the prototype for real world situations.

Figure 12 Development process of fabric pressure sensors into the sleeve garment and Pre and post voltage readings for the embedded textile-based pressure sensors in the wearable sleeve garment.

To showcase the benefit of the wearable µBrain sleeve prototype from an end user point of view, selected real-life use case scenarios were identified and detailed in [Fig 11(b)]. Through the selected placement and integration of the fabric pressure sensors into the sleeve garment, we focused on scenarios where these pressure points would be utilized in the everyday tasks of a human being. This application demonstrated that the wearable µBrain SNN prototype has the ability to act as a functioning wearable intelligent alert notification device for health scenarios where the human body has sensing limitations in their forearm.

Testing was completed and voltage output measurements obtained for each individual scenario. Slight fluctuations randomly occurred for expected outputs linked to the truth table inputs *010* and *101*. This again was mainly due to the set threshold level of *2.3 Volts* to differentiate between a logic '1' and logic '0'. It was observed during testing that the slight fluctuation in the obtained output voltage would increase or decrease slightly above or below this set threshold level and, hence, impact the desired output for these two truth table inputs. This issue can be rectified by altering the SMD weight resistor values in the neural network to ensure the desired output voltage reading range is not in close proximity to the set threshold level and, hence, any small fluctuations do not impact the end desired result.

### 4.3.2 Wearable µBrain stimulation of Acute Brain Slices

Inspired by the biological neural network of the human brain, the overriding concept envisions an application where the wearable µBrain acts as an intelligent artificial SNN fabric based wearable µBrain that has the potential to compute on-garment intelligence and relay signals from the wearable SNN back to the human brain (cerebrum) as shown in [Fig 13(a)(b)]. This vision would involve triggered outputs coming from the wearable µBrain connected via the spinal cord back to the brain's cerebrum for processing as shown in [Fig 13(c)]. This conceptual application envisions the use of the wearable µBrain specifically for health focused scenarios where the patient has a severe impediment that limits any forearm sensation or sense of awareness of the forearm location. Direct connection and relaying of information coming from the wearable µbrain sleeve application to a human brain for processing is not within the scope of this paper. Instead, we have focused initially on assessing the application and capability of the wearable µBrain to act as a trigger to stimulate acute brain slices using the wearable µBrain prototype and its SNN computational voltage output as a trigger for the stimulation.

### 4.3.2.1 Experimental protocol

A *4-months-old* FVB wild type mouse was culled by cervical dislocation and decapitated using scissors. The head was immediately immersed in semi-frozen *cutting* artificial cerebrospinal fluid (aCSF) [*$CaCl_2$ 0.5mM; glucose ($C_6H_{12}O_6$) 25mM; $MgSO_4$ 7mM; $NaH_2PO_4$ 1.25mM; KCl 2.5mM; $NaHCO_3$ 25mM; NaCl 87mM; sucrose ($C_{12}H_{22}O_{11}$) 75mM*] for a few seconds. The brain was then rapidly removed from the skull, using fine dissections tools, and placed in ice-cold *cutting* aCSF bubbled with *95% oxygen ($O_2$) / 5% carbon dioxide ($CO_2$),* while applying glue to a mounting block. The cerebellum and olfactory cortex were then removed using a razor blade, and the brain was cut along the midsagittal plane into two hemispheres. Both hemispheres were glued to the mounting block, with the parietal lobes facing upwards, and transferred to a vibratome bath (Campden Instruments *7000smz-2*) containing hyper-oxygenated ice-cold *cutting* aCSF. *400µm* thick parasagittal hippocampal slices were cut at a slow speed (*0.1mm/s*) and placed in a holding chamber with hyper-oxygenated *recording* aCSF [*$CaCl_2$ 2.5mM; glucose ($C_6H_{12}O_6$) 11mM; $MgSO_4$ 1mM; $NaH_2PO_4$ 1mM; KCl 2.5mM; $NaHCO_3$ 26.2mM; NaCl 119mM*] at room temperature (RT) for one hour. One hippocampal slice was subsequently transferred to a recording chamber and perfused with oxygenated *recording* aCSF (*95% $O_2$ / 5% $CO_2$*) at a stable speed of *3ml/min* and maintained at *35°C* for the duration of the recordings.

### 4.3.2.2 Experiment Execution

Hippocampal $CA_1$ pyramidal cells were identified using infrared DIC microscopy. The Shaffer-collateral commissural pathway was stimulated using a bipolar tungsten electrode placed approximately *50µm* from the cell body layer of the $CA_1$ hippocampal region. Evoked post-synaptic responses were recorded using a glass micropipette (Harvard Apparatus, Massachusetts, USA) filled with *recording* a CSF and placed *60-80mm* from the cell body layer. Multi-clamp *700B* amplifier interfaced with a Digidata 1550B and pCLAMP 11 software (all from Molecular Devices, CA, USA) were used. Events were low pass filtered at *2 kHz* and sampled at *10 kHz*.

To evoke synaptic responses from the brain slice, a "*111*" truth table input *5 Volts* trigger from the DC power supply was connected to the wearable µBrain garment, which was interfaced with the TTL input of a DS3 isolated current stimulator (Digitimer, Letchworth, UK). The output voltage of the wearable garment was ~3 *Volts*, and *3.2mA* current at *10 kHz* was delivered to the brain slice for approximately *11 secs*.

### 4.3.2.3 Experiment Results

Once this stimulation phase trigger occurred, we expected to see a trace showing an activation period ,there the trace measurements fluctuated and were recorded showing cell stimulation coming from the brain slice. A trace detailing the brain slice cell stimulation in the hippocampus regions (CA1) was obtained from the experiment [Fig 13(e)]. From the trace we can clearly see the initiation point when the stimulation commenced, the timeframe of *~11* seconds when we can see the brain slice being stimulated and then the disconnection of the wearable µBrain and ceasing of the stimulation of the brain slice. The experiment executed as expected , with the output results being the trace demonstrating the stimulation period of the brain slice when the wearable µBrain was connected.

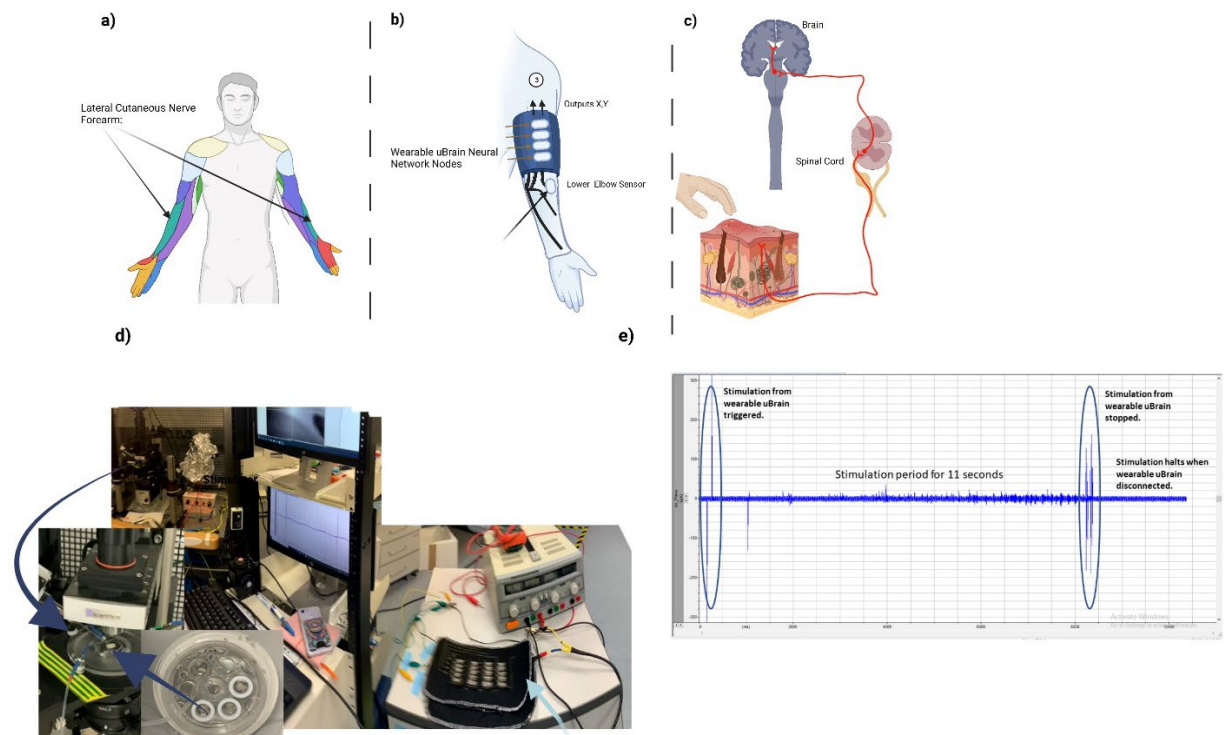

Figure 13 a) Identification of the location of the lateral Cutaneous Nerve in the forearm. b) Layout of the wearable µBrain on an arm sleeve with placement points for the sensors. c) Skin sensing capabilities through the spinal cord to be processed in the brain. d) Acute Brain Slice, stimulation through the wearable µBrain. e)Stimulation trace output result obatined folloing trigger from the wearable µBrain.

## 5   Discussion

ICT computational processing technologies and methods such as SNNs and neuromorphic computing are inspiring advancements in the next generation of E-textiles. The E-textiles and smart materials research domain is primed to take such technological advancements and maximize the potential impact through new innovations and applications The promise of E-textile based smart garments with embedded on-body intelligent neural network computing capabilities stem an exciting era especially within the healthcare domain. This paper focused on the design, development and validation of a wearable µBrain artificial SNN successfully incorporating key architectural design elements emerging from the neuromorphic driven research sector. The wearable µBrain with its modular, scalable and flexible properties can be applied across many sectors where textile based wearable applications exist. We successfully applied a SNN methodology to a textile-based fabric environment and validated its operational functionality linked to a focused healthcare use case. Our results show that such a prototype SNN demonstrator operates efficiently within a fabric-based environment.

The wearable µbrain offers a low size, weight and power (SWaP) solution that aligns with key architectural elements coming from the state-of-the-art neuromorphic processing vision. The overall size of the wearable µBrain architecture is reduced due to the use of SMD's as weights. A reduced size is also adopted through the modular multi-layered and scalable design approach adopted during implementation. From a textile application point of view this can greatly reduce the size of the wearable µBrain in a fabric environment. This was demonstrated via a wearable µBrain Sleeve garment. From a computational point of view, we have successfully implemented the capability to interchange weights and reduce or decrease the number of hidden layers in the wearable µBrain SNN. We conform to the low power requirement, as the wearable µBrain is an event driven SNN prototype. This enables the consumption of low power in comparison to other artificial neural network methods.

The overall conceptual vision to utilize the wearable µBrain to sense, compute and send triggered stimulation to the brain for processing was demonstrated through a focused use case where the wearable µBrain and its voltage output acting as a trigger, was incorporated into the protocol of stimulating acute brain slices.

Key findings of the study include

- Creation of a SNN/neuromorphic inspired architecture that enables bio-inspired on-garment computational intelligence in a fabric environment.
- Using off-the shelf easily obtainable SMD components to produce a wearable SNN with event driven artificial intelligence in a garment fabric environment. Development of a new method to add SMD devices to a fabric environment while also actively reducing the overall circuitry size.
- Design and development of a multilayer fabric hidden layer (FHL) neural network in a fabric environment.
    - A complex 3 input, 2 hidden layer, 2 output wearable SNN demonstrated through a wearable sleeve application connected to fabric pressure sensors in the garment.
    - The implementation of a multi-layered fabric design with hidden layers on separate fabric layers that can be connected / disconnected/ added to / reduced as required, enabling the prototype to be scalable, flexible and extendable.
- Implement an adaptive offline learning approach for the wearable µbrain SNN, through the use of interchangeable SMD resistor connector weights to the nodes in the artificial SNN.
- Creation of a wearable µBrain that meets current SWaP AI properties, low (size, weight and power).

Our proposed research drives the next generation of E-textiles, where textiles and bio inspired AI meet to create new and innovative SNN driven on-garment intelligence and computing research. This could usher in a new age of Internet of Bio-Nano Things for brain interface, where cyber systems are connected to neural cells for communications control (Bernal et al,.2021) (Balasubramaniam et al,. 2018). Future research will investigate the potential to develop a more enhanced wearable µBrain design, investigating further elements of SNN and neuromorphic technologies along with brain inspired methodologies to develop a next iteration prototype. Specific attention and further research will build on the current core elements of the wearable µBrain expanding out the research aspects linked to its robustness, fault tolerance, usability and performance. Also, further research will be investigated around the application of the wearable µBrain concept across other health focused use cases (for example stroke rehabilitation and epilepsy) , assessing its extendibility and potential to impact such critical conditions is a beneficial manner.

## 6  Acknowledgments

D. Henshall is funded in part by FutureNeuro from Science Foundation Ireland (SFI) under Grant Number 16/RC/3948 and co-funded under the European Regional Development Fund and by FutureNeuro industry partners.

## 7  References

A. L. et. al (2018), Loihi Asynchronous Neuromorphic Research Chip.24th IEEE International Symposium on Asynchronous Circuits and Systems (ASYNC) 2018 .DOI:10.1109/ASYNC.2018.00018.

An, M. A. Ehsan, Z. Zhou, F. Shen and Y. Yi.(2019). Monolithic 3D neuromorphic computing system with hybrid CMOS and memristor-based synapses and neurons. Integration 2019 Vol. 65 Pages 273-281.DOI: 10.1016/j.vlsi.2017.10.009.

Balaji (2021). Effects of Multiple Sclerosis on Motor Movement.Journal of Multiple Sclerosis 2021 Vol. 8 Issue 3 Pages 1-1. DOI: 10.35248/2376-0389.21.8.237.

Balaji, Adarsha, et al. (2019) "Mapping spiking neural networks to neuromorphic hardware." *IEEE Transactions on Very Large Scale Integration (VLSI) Systems* 28.1) 76-86.DOI: 10.1109/TVLSI.2019.2951493.

Balasubramaniam, S., Wirdatmadja, S., Barros, M. T., Koucheryavy, Y., Stachowiak, M., Jornet, J. M.. (2018) Wireless Communications for Optogenetics-Based Brain Stimulation: Present Technology and Future Challenges. IEEE Communications Magazine Vol. 56 Issue 7 Pages 218-224 July. DOI: 10.1109/MCOM.2018.1700917.

Bernal, S. L. Celdrán, A. H., Pérez, G. M., Barros, M. T., Balasubramaniam, S.,  (2021) Security in Brain-Computer Interfaces: State-of-the-art, Opportunities, and Future Challenges. ACM Computing Survey Vol. 54 Issue  1. DOI:10.1145/3427376.

Burkitt AN (2006). A review of the integrate-and-fire neuron model: II. Inhomogeneous synaptic input and network properties. Biol Cybern, DOI: 10.1007/s00422-006-0082-8.

Covi, E., Donati, E., Liang, X., Kappel, D., Heidari, H., Payvand, M., & Wang, W. (2021). Adaptive Extreme Edge Computing for Wearable Devices. *Frontiers in neuroscience*, *15*, 611300. DOI: 10.3389/fnins.2021.611300.

Cheng, H., Wen, W., Wu, C., Li, S., Li, H.H., & Chen, Y. (2017). Understanding the design of IBM neurosynaptic system and its tradeoffs: A user perspective. *Design, Automation & Test in Europe Conference & Exhibition (DATE), 2017*, 139-144.DOI: 10.23919/DATE.2017.7926972

Davies *et al*., "Loihi: A Neuromorphic Manycore Processor with On-Chip Learning," in *IEEE Micro*, vol. 38, no. 1, pp. 82-99, January/February 2018, DOI: 10.1109/MM.2018.112130359.

Fouda, Mohammed E., Ahmed M. Eltawil, and Fadi Kurdahi. (2019) "On resistive memories: One step row readout technique and sensing circuitry." *arXiv preprint arXiv:1903.01512* .

Garain, A., Basu, A., Giampaolo, F. et al. (2021) Detection of COVID-19 from CT scan images: A spiking neural network-based approach. Neural Comput & Applic 33, 12591–12604. DOI:https://doi.org/10.1007/s00521-021-05910-1.

Ham, M. Kang, S. Jang, J. Jang, S. Choi, T.-W. Kim, et al. One-dimensional organic artificial multi-synapses enabling electronic textile neural network for wearable neuromorphic applications.Science Advances 2020 Vol. 6 Issue 28 Pages eaba1178 DOI:10.1126/sciadv.aba1178.

Herculano-Houzel, S. (2009). "The human brain in numbers: a linearly scaled-up primate brain." Frontiers in Human Neuroscience 2009 Vol. 3. DOI: 10.3389/neuro.09.031.2009.

Herzfeld, D. J. and S. A. Beardsley (2011). Synaptic weighting for physiological responses in recurrent spiking neural networks. 2011 Annual International Conference of the IEEE Engineering in Medicine and Biology Society, IEEE. DOI: 10.1109/IEMBS.2011.6091039. PMID: 22255262.

Indiveri, G. (2021). Introducing 'Neuromorphic Computing and Engineering'. *Neuromorphic Computing and Engineering*, *1*(1), 010401.

Innocenti, G., et al. (2021). "Memristor Circuits for Simulating Neuron Spiking and Burst Phenomena. Frontiers in Neuroscience 2021 Vol. 15. DOI: 10.3389/fnins.2021.681035.

Jang, J. Lee and S. S. Yeo (2021). Central post-stroke pain due to injury of the spinothalamic tract in patients with cerebral infarction: a diffusion tensor tractography imaging study
Neural regeneration research 2017 Vol. 12 Issue 12 Pages 2021. DOI**:** 10.4103/1673-5374.221159

Jayaraman, M. (2020). "Applications of Neuromorphic Computing." https://docs.lib.purdue.edu/cgi/viewcontent.cgi?article=1010&context=ideas .

Ji, A.-N. Cha, S.-A. Lee, S. Bae, S. H. Lee, D. S. Lee, et al.(2016). Integrated all-organic 8× 8 one transistor-one resistor (1T-1R) crossbar resistive switching memory array.Organic Electronics 2016 Vol. 29 Pages 66-71.

Kadir, L., et al. (2018). "Emerging Roles of the Membrane Potential: Action Beyond the Action Potential." Frontiers in Physiology, DOI: 10.3389/fphys.2018.01661.

Ker, J., Wang, L., Rao, J., & Lim, T. (2017). Deep learning applications in medical image analysis. *Ieee Access*, *6*, 9375-9389. DOI**:** 10.1109/ACCESS.2017.2788044

Koyanagi, Y. Nakagawa, K.-W. Lee, T. Nakamura, Y. Yamada, K. Inamura, et al. (2001). Neuromorphic vision chip fabricated using three-dimensional integration technology 2001 IEEE International Solid-State Circuits Conference. Digest of Technical Papers. ISSCC (Cat. No. 01CH37177) 2001. Publisher : IEEE Pages : 270-271.

Lapicque, L. (1907). "Louis Lapicque." J. physiol **9**: 620-635.

Lee, V. K. Sangwan, W. A. G. Rojas, H. Bergeron, H. Y. Jeong, J. Yuan, et al. (2020) Dual-Gated MoS2 Memtransistor Crossbar Array. Advanced Functional Materials Vol. 30 Issue 45 Pages 2003683.DOI: 10.1002/adfm.202003683.

Liao, F., et al. (2021) "Neuromorphic vision sensors: Principle, progress and perspectives." Journal of Semiconductors. DOI: 10.1088/1674-4926/42/1/013105.

Mehonic, A., et al. (2020). "Memristors—From In-Memory Computing, Deep Learning Acceleration, and Spiking Neural Networks to the Future of Neuromorphic and Bio-Inspired Computing." Advanced Intelligent Systems . DOI: 10.1002/aisy.202000085.

Mokhlespour Esfahani, M. I., and Maury A Nussbaum (2019). "Classifying Diverse Physical Activities Using "Smart Garments." Sensors (Basel, Switzerland) DOI:10.3390/s19143133.

Molla, M. T. I., et al. (2017). Surface-mount manufacturing for e-textile circuits. Proceedings of the 2017 ACM International Symposium on Wearable Computers. Maui, Hawaii, Association for Computing Machinery**:** 18–25.DOI: 10.1145/3123021.3123058.

Pang, L., et al. (2020). "Case Study—Spiking Neural Network Hardware System for Structural Health Monitoring." *Sensors* 2020, *20*, 5126.DOI: 10.3390/s20185126.

Park, H.-L., et al. (2020). "Flexible Neuromorphic Electronics for Computing, Soft Robotics, and Neuroprosthetics." Advanced Materials . DOI: 10.1002/adma.201903558.

Park, J. (2020). "Neuromorphic Computing Using Emerging Synaptic Devices: A Retrospective Summary and an Outlook." Electronics.  DOI: 10.3390/electronics9091414.

Pfeiffer, M. and T. Pfeil (2018). "Deep Learning With Spiking Neurons: Opportunities and Challenges." Frontiers in Neuroscience. DOI: 10.3389/fnins.2018.00774.

Rajan, K., et al. (2018). "Wearable Intrinsically Soft, Stretchable, Flexible Devices for Memories and Computing." *Sensors (Basel, Switzerland)*, *18*(2), 367.DOI: 10.3390/s18020367.

Ser, J.D., Osaba, E., Molina, D., Yang, X., Salcedo-Sanz, S., Camacho, D., Das, S., Suganthan, P.N., Coello, C.A., & Herrera, F. (2019). Bio-inspired computation: Where we stand and what's next. *Swarm Evol. Comput., 48*, 220-250. DOI: 10.1016/j.swevo.2019.04.008.

Seufert, M. HassanpourAmiri, P. Gkoupidenis and K. Asadi (2021). Crossbar Array of Artificial Synapses Based on Ferroelectric Diodes. Advanced Electronic Materials 2021 Pages 2100558.DOI: 10.1002/aelm.202100558.

Shahid, Nida et al. (2019)  "Applications of artificial neural networks in health care organizational decision-making: A scoping review." *PloS one* vol. 14,2 e0212356. 19 Feb. 2019. DOI:10.1371/journal.pone.0212356.

Stuijt, J., et al. (2021). "µBrain: An Event-Driven and Fully Synthesizable Architecture for Spiking Neural Networks." Frontiers in Neuroscience. DOI: 10.3389/fnins.2021.664208.

Sun and D. Wen (2020). Multistage resistive switching behavior organic coating films-based of memory devices. Progress in Organic Coatings 2020 Vol. 142 Pages 105613.DOI: 10.1016/j.porgcoat.2020.105613.

Tuchman, Y., et al. (2020). "Organic neuromorphic devices: Past, present, and future challenges." MRS Bulletin MRS Bulletin, 45(8), 619-630. DOI:10.1557/mrs.2020.196.

Wan, H., et al. (2020). "Flexible carbon nanotube synaptic transistor for neurological electronic skin applications." ACS nano DOI:10.1021/acsnano.0c04259.

C.-H. Wu and M. Boudier-Revéret. (2019). Ultrasound-guided steroid injections for lateral antebrachial cutaneous nerve entrapment within postsurgical scar.American journal of physical medicine & rehabilitation 2019 Vol. 98 Issue 9 Pages e106. DOI: 10.1097/PHM.0000000000001150.

Xia, Q., Yang, J.J. (2019) Memristive crossbar arrays for brain-inspired computing. *Nat. Mater.* 18, 309–323.DOI: 10.1038/s41563-019-0291-x.

Yao, P., Wu, H., Gao, B. et al. (2020) Fully hardware-implemented memristor convolutional neural network. Nature 577, 641–646. DOI: 10.1038/s41586-020-1942-4.

Yang, J. Wang, B. Deng, C. Liu, H. Li, C. Fietkiewicz, et al. (2018) Real-time neuromorphic system for large-scale conductance-based spiking neural networks. IEEE transactions on cybernetics 2018 Vol. 49 Issue 7 Pages 2490-2503.  DOI: 10.1109/TCYB.2018.2823730.

Yuan, S. E. Liu, A. Shylendra, W. A. Gaviria Rojas, S. Guo, H. Bergeron, et al. (2021). Reconfigurable MoS2 Memtransistors for Continuous Learning in Spiking Neural Networks. Nano letters 2021 Vol. 21 Issue 15 Pages 6432-6440. DOI: 10.1021/acs.nanolett.1c00982.
24